 \documentclass[final,5p,times,twocolumn]{elsarticle}

\usepackage{amssymb}
\usepackage{lipsum}
\usepackage{xcolor}
\usepackage{url} 
\usepackage{amsmath}
\usepackage{tabularx}
\definecolor{pink}{rgb}{1,0.75,0.8}

\journal{}

\begin{document}

\begin{frontmatter}



\title{AI-Driven Scenarios for Urban Mobility: Quantifying the Role of ODE Models and Scenario Planning in Reducing Traffic Congestion.}

\author[first]{Katsiaryna Bahamazava}
\affiliation[first]{organization={Department of Mathematical Sciences G.L. Lagrange, Politecnico di Torino},
            addressline={Corso Duca degli Abruzzi, 24}, 
            city={Torino},
            postcode={10129}, 
            country={Italy}}

\begin{abstract}
\indent Urbanization and technological advancements are reshaping urban mobility, presenting both challenges and opportunities. This paper investigates how Artificial Intelligence (AI)-driven technologies can impact traffic congestion dynamics and explores their potential to enhance transportation systems' efficiency. Specifically, we assess the role of AI innovations, such as autonomous vehicles and intelligent traffic management, in mitigating congestion under varying regulatory frameworks. Autonomous vehicles reduce congestion through optimized traffic flow, real-time route adjustments, and decreased human errors.

The study employs Ordinary Differential Equations (ODEs) to model the dynamic relationship between AI adoption rates and traffic congestion, capturing systemic feedback loops. Quantitative outputs include threshold levels of AI adoption needed to achieve significant congestion reduction, while qualitative insights stem from scenario planning exploring regulatory and societal conditions. This dual-method approach offers actionable strategies for policymakers to create efficient, sustainable, and equitable urban transportation systems. While safety implications of AI are acknowledged, this study primarily focuses on congestion reduction dynamics.

\end{abstract}




\end{frontmatter}



\section{Introduction}
Urbanization is accelerating globally, with more than half of the world's population now living in urban areas—a figure projected to rise to nearly 70\% by 2050. This rapid urban growth intensifies challenges related 
to urban mobility, including traffic congestion, environmental pollution, and inefficiencies in 
transportation systems. Traditional solutions, such as expanding infrastructure or implementing policy interventions like congestion pricing, have often proved insufficient or introduced new complexities \cite{cheng2024interventions}. As cities strive to become more livable and sustainable, there is a pressing need for innovative, forward-thinking approaches that integrate a suite of modern technologies.

Foresight and scenario planning have emerged as critical tools in addressing these complex challenges. By anticipating future trends and uncertainties, these 
methodologies enable policymakers and business leaders to develop strategies that are robust under various 
possible futures. In the context of urban mobility, foresight practices help to undertand how emerging 
technologies, societal shifts, and regulatory changes may interact to shape transportation systems over the coming decades.
Among these modern technologies, Artificial Intelligence (AI) stands at the forefront, 
holding significant promise to revolutionize urban mobility. Yet, AI does not operate in isolation. It is 
increasingly complemented by an ecosystem of digital innovations: Internet of Things (IoT) sensors that 
collect real-time traffic data, 5G and next-generation communication networks that ensure rapid data transfer, 
blockchain platforms that secure transactions and enhance trust, and cloud computing infrastructures that enable scalable data analytics 
\cite{bijalwan2024navigating}. These technologies, working in tandem, can enhance predictive analytics, improve system interoperability, and ultimately drive more adaptive and responsive transportation services. 
For example, AI-driven autonomous vehicles, when combined with IoT-enabled traffic lights and 5G connectivity, can synchronize traffic flows more efficiently; similarly, secure data exchange through blockchain can improve service reliability and foster user confidence in shared mobility models \cite{ahmed2024towards}, \cite{luusua2023urban}. 
However, the integration of AI, alongside other emerging digital tools, also presents challenges, 
including ethical considerations, regulatory hurdles, and the risk of exacerbating social inequalities \cite{rahman2023impacts}. Understanding how these 
complementary technologies interact, evolve, and influence one another is critical to forging sustainable urban mobility solutions that are both technologically feasible and socially equitable.

This paper integrates foresight and scenario planning with mathematical modeling to explore how AI-driven technologies can reshape urban transportation systems. Specifically, we employ Ordinary Differential Equations (ODEs) to model the interaction between AI adoption and traffic congestion over time. 
 In selecting an appropriate modeling approach, ODEs provide a balance between complexity and interpretability. Unlike more granular agent-based models that simulate individual actors or discrete-
event simulations that often demand extensive computational resources, ODEs allow for capturing the 
aggregate dynamics of AI adoption and traffic congestion in a continuous and mathematically tractable manner. This aggregated, system-level perspective is well-suited to the early-stage strategic planning and 
scenario analysis employed in this study. Additionally, ODE models facilitate sensitivity analyses and parameter 
estimation, making it easier to examine how different 
regulatory conditions or technological adoption rates influence outcomes. While 
more complex models might offer finer detail, ODEs 
prove sufficient for our aim of understanding broad, long-term trends and identifying critical thresholds that inform policy and regulatory strategies.
These ODE models allow us to simulate and quantify the effects of different regulatory and societal conditions on the adoption of AI 
technologies and the resulting mobility outcomes. By using Python for 
numerical simulations, we analyze the dynamics of urban mobility across multiple future scenarios, offering both qualitative and quantitative insights.

The structure of the paper is as follows:

Section~\ref{sec:Literature-Review} reviews the current challenges in urban mobility, the limitations of traditional solutions, and the role of AI technologies, with a focus on foresight and scenario planning methodologies.

Section~\ref{sec:Methodology} outlines the research methods used, 
including scenario development, the ODE-based modeling approach, and Python simulations.

Section~\ref{sec:Modelling} presents detailed future 
scenarios of urban mobility influenced by AI, 
considering factors such as regulatory changes, 
technological advancements, and societal trends, along with simulation results.

Section~\ref{sec:Discussion} analyzes the implications of these scenarios for policymakers, industry stakeholders, and urban residents.

Section~\ref{sec:Conclusion} concludes the paper with strategic recommendations and insights into the future of AI in urban transportation, emphasizing the 
potential of combining foresight methods with mathematical modeling.

By applying both foresight and ODE-based modeling, this research offers a comprehensive approach to understanding and shaping the future of AI-driven urban mobility. Through scenario planning and quantitative simulations, we provide practical guidance for stakeholders seeking to navigate the complexities of integrating AI into urban transportation systems.

\section{Literature Review}
\label{sec:Literature-Review}
Urban mobility faces multifaceted challenges, including traffic 
congestion, environmental pollution, inefficiencies in public 
transportation, and the pressures of increasing urbanization. 
Traditional solutions have often proved inadequate, prompting a shift 
towards innovative technologies like Artificial Intelligence (AI) to address these issues comprehensively.

Traffic congestion significantly impacts economies both directly and indirectly. \cite{sweet2011does} demonstrates that congestion leads to immediate costs such as travel delays and unreliable travel times, which reduce productivity as workers and goods take longer to reach their destinations. While travelers may adapt by altering routes, travel times, or modes of transportation, the overall economic costs remain substantial. Moreover, congestion influences broader economic geographies by affecting where businesses and residents choose to locate, often redistributing rather than eliminating economic growth. Public-sector efforts to mitigate congestion through road expansion have been costly and largely ineffective, leading experts to advocate for adaptation strategies over mitigation \cite{sweet2011does}.

Environmental pollution is another critical issue linked to transportation systems. \cite{jacyna2017noise} highlight that transportation, particularly road transport, is a major contributor to environmental pollution through emissions of harmful compounds like carbon dioxide (CO$_2$), nitrogen oxides (NO$_x$), particulate matter (PM), and hydrocarbons (HC). These pollutants adversely affect human 
health and the environment. Additionally, noise pollution from transportation has become a serious public health concern, causing sleep disturbances, cardiovascular diseases, and stress among urban populations. Efforts to mitigate these negative effects focus on reducing emissions and noise levels through technological advancements in vehicles, traffic management strategies, and the development of sustainable transportation systems \citep{jacyna2017noise}.

Inefficiencies in public transportation systems exacerbate urban mobility challenges. \cite{boile2001estimating} identifies technical and scale inefficiencies as primary issues affecting transit systems' performance and resource utilization. Technical inefficiencies arise from suboptimal use of resources such as labor, fuel, and maintenance, 
while scale inefficiencies occur when the size of the transit operation leads to overcapacity or underutilization. Using data envelopment analysis (DEA), studies have assessed and compared the efficiency of various transit systems, pinpointing areas where improvements can optimize operations. Addressing these inefficiencies can enhance service delivery, reduce operational costs, and increase the sustainability of urban transportation networks \cite{boile2001estimating}.

The rapid pace of urbanization and increasing population density significantly impact energy consumption and efficiency in cities, particularly concerning transportation systems. \cite{he2023city} note that urbanization concentrates population and economic activities, escalating the demand for transportation and energy. While densely populated areas can benefit from the agglomeration effect—where resources are used more efficiently due to proximity—overpopulation can lead to congestion and inefficiencies. Cities with inadequate 
transportation infrastructure face greater challenges, as increased demand for travel between dispersed urban sub-centers results in longer commuting times, more vehicle use, and higher energy consumption. Efficient public transportation systems in well-planned urban areas can mitigate these negative effects, improving energy efficiency and reducing pollution. Thus, urban planning strategies that balance density with adequate infrastructure are crucial in addressing the challenges posed by increasing urbanization \cite{he2023city}.

Traditional policy interventions like congestion pricing have been implemented to reduce traffic congestion but possess inherent limitations. \citet{cheng2024interventions} discuss how congestion pricing, which involves charging drivers for road usage during peak times, aims to decrease the number of vehicles on the road and alleviate congestion. While cities like London, Stockholm, and 
Singapore have reported significant traffic reductions when congestion pricing is paired with public transport improvements, these policies can have unintended consequences. Traffic may shift to non-charged periods or alternative routes outside the charging zone, and wealthier drivers may continue to use the roads while lower-income individuals face restricted access. The success of congestion pricing largely depends on complementary measures, such as enhanced public transit options, to sustain long-term benefits \cite{cheng2024interventions}.

In response to these challenges, AI technologies have emerged as 
promising tools to revolutionize urban mobility, particularly within 
the context of smart city development. \cite{luusua2023urban} explore how AI-powered systems are integrated into various aspects of transportation, including autonomous vehicles, traffic management, and 
urban planning. Adaptive algorithms optimize routes, manage congestion, and enhance public transportation efficiency. Technologies like computer vision, machine learning, and AI-driven analytics analyze data from urban mobility systems, geographic information systems (GIS), and surveillance networks to create sophisticated models of city traffic and behavior. AI applications in autonomous driving and traffic automation are pivotal for the future of urban mobility, promising to reduce human error and streamline transportation systems. However, implementing these technologies raises ethical questions regarding privacy, surveillance, and accessibility, necessitating thoughtful and inclusive approaches to integrating AI into urban mobility systems \cite{luusua2023urban}.

Autonomous vehicles (AVs) represent a significant advancement in AI applications within urban mobility. \cite{rahman2023impacts} explain that AVs rely on advanced sensor technologies such as LiDAR, radar, and cameras to perceive their environment, feeding data into AI algorithms that enable real-time decision-making. These systems detect objects like pedestrians, cyclists, and other vehicles, predict their movements, and make critical navigation decisions. AVs hold great promise for reducing accidents by eliminating many human errors responsible for traffic collisions. However, several challenges impede their widespread adoption. Legal and regulatory frameworks surrounding autonomous driving are underdeveloped, raising questions about liability in accidents. Ethical concerns arise, particularly regarding decision-making in unavoidable accident scenarios. Additionally, public acceptance and trust remain significant hurdles, as many people are skeptical about the safety and reliability of fully autonomous systems. Overcoming these regulatory, ethical, and societal challenges is crucial for AVs to effectively transform urban mobility \cite{rahman2023impacts}.

Adaptive traffic signals powered by AI algorithms are revolutionizing traffic management by improving flow and reducing congestion. \cite{musa2023sustainable} discuss how these systems utilize data from sensors and cameras to monitor vehicle patterns and dynamically adjust signal timings in response to real-time traffic conditions. By optimizing traffic signals, cities can reduce idle times, lower emissions, and enhance the overall efficiency of transportation networks. AI-driven systems like the Sydney Coordinated Adaptive Traffic System (SCATS) and the Split, Cycle, and Offset Optimization Technique (SCOOT) have successfully managed traffic congestion in cities worldwide. However, challenges such as developing appropriate legal frameworks, addressing ethical concerns, and building public trust remain key barriers to the widespread adoption of AI in traffic management \citep{musa2023sustainable}.

AI technologies also play a crucial role in public transportation by optimizing routes and schedules and enhancing the passenger experience. \cite{jevinger2024artificial} highlight how AI algorithms analyze extensive datasets, including traffic patterns and passenger demand, to create demand-responsive transit systems that adapt to real-time conditions. This approach helps predict passenger flow and improve resource allocation, ensuring that transit services meet varying demands throughout the day. AI-powered assistants and applications provide passengers with real-time notifications regarding arrival times, delays, or route changes, and offer personalized services such as optimal route suggestions based on individual preferences. Additionally, AI facilitates predictive maintenance by analyzing vehicle data to anticipate and prevent potential failures, improving fleet management and reducing downtime. These advancements contribute to a more efficient, reliable, and passenger-friendly public transportation system \cite{jevinger2024artificial}.

While a growing body of literature examines AI applications in autonomous vehicles (in addition to discussed above see e.g., \cite{garikapati2024autonomous}, \cite{singh2024autonomous}), intelligent traffic management systems (e.g., \cite{alanazi2024interoperability}), and demand-responsive transit (\cite{zheng2024optimizing}), several gaps remain. For instance, most studies emphasize either technical feasibility or policy frameworks in isolation, rarely integrating both to provide a holistic view of how regulatory conditions influence AI adoption and its subsequent impact on congestion reduction.
Another underexplored area concerns the quantitative assessment of these AI-driven transformations under varying regulatory and societal scenarios. While scenario planning methodologies are sometimes employed qualitatively, there is limited empirical work using mathematical models to simulate how different adoption rates, policy environments, and public attitudes might interact to influence outcomes. Few studies have combined foresight methods and rigorous quantitative modeling to identify thresholds of AI adoption that significantly reduce congestion, or to determine how regulatory support could accelerate or hinder these advancements.
By directly addressing these gaps, the present study contributes to the literature in the following ways. Our paper incorporates a scenario-based ODE modeling framework to quantitatively explore how 
various regulatory and societal factors influence the long-term evolution of AI adoption and congestion 
dynamics. In addition, the study provides actionable insights for policymakers and practitioners, identifying critical adoption thresholds and regulatory levers that may guide sustainable, data-driven strategies for future urban transportation planning.

\section{Methodology} \label{sec:Methodology} 

This study employs a foresight and scenario planning approach to explore how AI-driven technologies may shape the future of urban mobility. The methodology integrates scenario development with modelling 
techniques, providing both qualitative and quantitative insights into potential futures. This combination strengthens the robustness of the analysis, ensuring that findings are applicable to real-world strategic planning.

\subsection{Scenario Development}

Scenario development is central to this research, allowing us to explore multiple plausible futures of 
urban mobility influenced by AI technologies. The use of scenario planning helps stakeholders to prepare for a range of potential outcomes, considering both known trends and emerging uncertainties.


\subsubsection{Identifying and Validating Key Drivers and Uncertainties}

The utilization of AI in urban mobility systems, particularly for safety-critical applications such as autonomous vehicles and intelligent traffic management, hinges on two interrelated factors: the rate of AI adoption and the degree of regulatory support. These drivers were identified through an extensive review of literature (e.g., \cite{luusua2023urban}, \cite{rahman2023impacts}, \cite{musa2023sustainable}, \cite{perez2024ai}) addressing the technical, societal, and regulatory challenges associated with AI deployment, alongside preliminary analysis of pilot AI-transportation projects \cite{eit2021urban}. These steps reinforced our initial assumption that variations in AI adoption and regulatory involvement are both highly impactful and uncertain. We thus selected these factors as axes for our scenario matrix.

Pilot AI projects provide valuable insights into the practicalities and challenges of urban mobility transformations. For instance, the \textit{AI-TraWell} project highlights how AI-powered travel assistants can promote personalized, sustainable travel by integrating user feedback and mobility service data to optimize travel routes based on individual preferences. Similarly, \textit{Siemens Mobility}'s deployment in Hagen, Germany, optimized traffic light control, reducing intersection waiting times by up to 47\%, demonstrating the role of AI in addressing urban congestion. In Niepołomice, Poland, the \textit{Tele-Bus} on-demand bus system illustrates how AI can dynamically manage flexible transportation services, reducing emissions and improving accessibility in low-density areas \cite{eit2021urban}.

Micromobility innovations such as \textit{Spin Insight Level 2 e-scooters}, which use AI to detect sidewalks and provide real-time feedback to riders, reflect AI's potential in improving safety and compliance in urban mobility. Additionally, autonomous mobility pilots such as Finland’s \textit{GACHA driverless bus} and Frankfurt Airport’s AI prediction models for flight arrivals further emphasize the interplay between technological advancements and regulatory environments. These projects collectively highlight the importance of AI adoption and regulatory frameworks as critical drivers in shaping urban mobility systems.

\textit{AI Adoption} is a dynamic and multifaceted process influenced by technological readiness, societal trust, and operational viability. Existing research highlights the fragmented nature of AI deployment in safety-critical domains, such as transportation, where significant disparities exist in the pace of adoption across regions and industries \cite{perez2024ai}. For instance, the integration of AI systems is often constrained by the need for robust performance under uncertain real-world conditions, explainability of decision-making processes, and compliance with safety assurance frameworks.

Trust in AI systems plays a pivotal role in driving adoption. Factors such as explainability, provability, and robustness are critical to building public and industry confidence in AI-driven technologies \cite{perez2024ai}. Moreover, the interplay between technical advancements and societal acceptance of these technologies directly impacts their adoption trajectory. For urban mobility, the effectiveness of AI in addressing congestion, pollution, and inefficiencies is contingent upon widespread acceptance and integration into existing transportation ecosystems.

The second critical driver, \textit{Regulatory Support}, encompasses the creation and enforcement of standards, guidelines, and policies that enable the safe and effective integration of AI technologies. The importance of regulatory support is highlighted in evolving AI-specific safety standards, such as VDE-AR-E 2842-61 and ISO 21448, which aim to ensure the safe integration of AI systems in safety-critical contexts \cite{perez2024ai}. These standards emphasize the need for comprehensive testing, safety assurance, and explainability of AI models.

Regulatory frameworks also play a central role in mitigating risks associated with AI deployment. The literature underscores that inadequate or inconsistent regulation can lead to safety failures, ethical concerns, and reduced public trust, all of which hinder the adoption of AI in critical applications such as urban mobility \cite{perez2024ai}. Conversely, clear and supportive policies can provide the necessary infrastructure and incentives for adopting AI technologies while addressing societal concerns such as data privacy, cybersecurity, and fairness.

In the context of urban mobility, the interplay between regulatory support and AI adoption is particularly salient. Strong regulatory frameworks not only facilitate the deployment of AI-driven technologies but also ensure that their implementation aligns with broader societal goals, such as sustainability and equity. This dual role of regulation— as both an enabler and a safeguard—reinforces its importance as a critical and uncertain driver in shaping the future of AI in transportation systems.

\subsubsection{Scenario Framework and Development}

Once key drivers and uncertainties are identified, a two-axis scenario planning method is employed. The two most critical and uncertain drivers—such as the degree of AI adoption and regulatory support—are used to develop a matrix of four distinct scenarios. This framework helps explore the range of plausible futures:

\begin{itemize}
    \item High AI Adoption with Strong Regulatory Support
    \item High AI Adoption with Weak Regulatory Support
    \item Low AI Adoption with Strong Regulatory Support
    \item Low AI Adoption with Weak Regulatory Support
\end{itemize}

Each scenario presents a unique combination of regulatory, societal, and technological factors shaping urban mobility.

\subsubsection{Scenario Narratives}

For each scenario, detailed narratives are developed to explore the implications for urban mobility. These narratives include:

\begin{itemize}
    \item Technological Integration: The extent of AI deployment in transportation systems, including autonomous vehicles, traffic management, and public transportation.
    \item Regulatory Environment: How supportive or restrictive regulations may impact the implementation of AI in urban mobility.
    \item Public Acceptance: Societal attitudes towards AI-driven changes in transportation and the resulting shifts in mobility behavior.
    \item Environmental Impact: The potential effects on sustainability and emissions reduction in each scenario.
\end{itemize}

\begin{figure}[htp]
    \centering
    \includegraphics[width=0.5\textwidth]{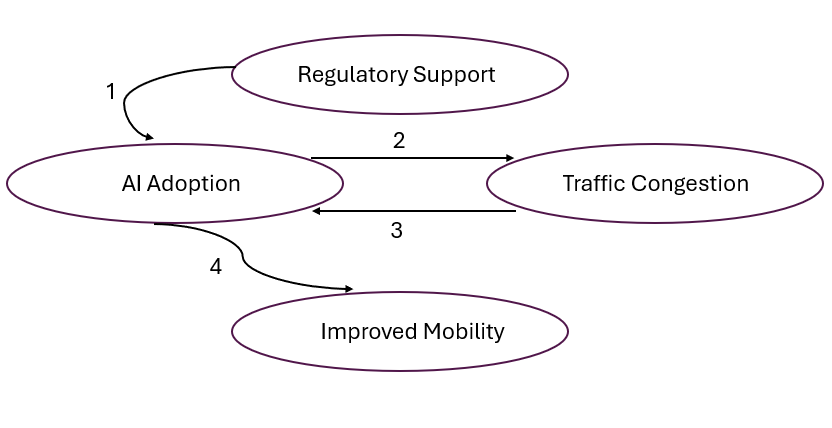}
    \caption{Interactions Between Regulatory Support, AI Adoption, and Traffic Congestion.}
\end{figure}

Figure 1 shows the dynamic interactions between AI adoption, traffic congestion, and regulatory support in urban mobility systems. The relationships are illustrated as follows:
\begin{enumerate}
\item Regulatory Support → AI Adoption: Facilitates adoption.
\item AI Adoption → Traffic Congestion: Reduces congestion.
\item Traffic Congestion → AI Adoption: Provides negative feedback, slowing adoption.
\item AI Adoption → Improved Mobility: Improves overall mobility and reduces congestion.
\end{enumerate}
\paragraph{Scenario 1: High AI Adoption with Strong Regulatory Support (Optimistic Future)}

\textbf{Technological Integration:}  
In this scenario, AI technologies are deeply integrated into urban mobility systems. Autonomous vehicles (AVs) dominate personal and commercial transportation, leading to the widespread use of AI-powered public transportation, smart traffic management systems, and Mobility-as-a-Service (MaaS) platforms. AI-driven solutions like predictive traffic management, dynamic congestion pricing, and personalized travel services ensure seamless, efficient, and user-friendly urban transportation.

\textbf{Regulatory Environment:}  
Governments actively support AI innovation in transportation, providing clear regulatory frameworks and incentivizing the development and deployment of AI technologies. International standards for AV safety, data sharing, and ethical AI usage are in place, ensuring both consumer protection and rapid technological advancement. Collaborative public-private partnerships drive forward AI adoption while maintaining stringent data privacy and cybersecurity regulations.

\textbf{Public Acceptance:}  
Society is generally optimistic and open to AI-driven changes in transportation. Public trust is high due to transparent government policies, safety protocols, and ethical AI practices. Citizens adapt to a shared-mobility model, reducing car ownership in favor of shared autonomous vehicles and efficient public transport systems. AI applications improve commuting experiences, with real-time travel updates and personalized services.

\textbf{Environmental Impact:}  
Strong regulatory support results in policies that prioritize sustainability alongside technological advancement. Cities shift to green energy sources, such as electrification powered by renewable energy, to support AV fleets and public transport. AI optimizes traffic flow and reduces congestion, leading to significant decreases in carbon emissions and energy consumption. This scenario represents the ideal convergence of AI-driven efficiency and sustainability.

\paragraph{Scenario 2: High AI Adoption with Weak Regulatory Support (Tech-Driven but Unregulated Future)}

\textbf{Technological Integration:}  
AI technologies are rapidly adopted across urban transportation systems, but without robust regulatory oversight. Autonomous vehicles, AI-powered logistics, and advanced traffic management systems are widely used, particularly in tech-driven cities. However, the lack of standardized safety regulations and data governance results in uneven technological integration across different urban centers.

\textbf{Regulatory Environment:}  
The absence of strong regulatory frameworks leads to fragmented AI adoption. While tech companies drive innovation and deploy cutting-edge AI solutions, there is little government oversight or intervention. As a result, key concerns such as data privacy, AI bias, and security remain inadequately addressed. Regulatory uncertainty leads to varying levels of safety standards for autonomous vehicles, creating a patchwork of rules across cities and countries.

\textbf{Public Acceptance:}  
Public trust in AI transportation technologies is mixed. Early adopters and tech-savvy consumers embrace the benefits of autonomous vehicles and AI-driven services, while others remain skeptical due to safety incidents and privacy concerns. A lack of cohesive regulations and transparency erodes trust, resulting in inequitable access to AI-driven services, with wealthier regions enjoying more reliable and safer systems than poorer ones.

\textbf{Environmental Impact:}  
Without clear regulatory mandates, AI’s potential for environmental impact reduction is only partially realized. While some cities see reduced emissions due to AI-optimized traffic and electric autonomous fleets, other regions continue to rely on fossil-fuel-powered transport due to a lack of coordinated efforts toward sustainable energy integration. This results in inconsistent environmental outcomes, with some cities reaping the benefits of AI-driven sustainability while others lag behind.

\paragraph{Scenario 3: Low AI Adoption with Strong Regulatory Support (Regulation-Driven Future with Limited Technology)}

\textbf{Technological Integration:}  
In this scenario, AI adoption in transportation is limited, despite strong government support. Regulatory frameworks are in place to encourage AI innovation, but technological advancement in urban mobility is slow due to high development costs, technical challenges, or public resistance. Autonomous vehicles and AI-powered systems exist but remain in pilot phases or are deployed in niche markets such as logistics or public transit.

\textbf{Regulatory Environment:}  
Governments play a key role in shaping the regulatory landscape, ensuring that any AI technologies deployed in transportation adhere to strict safety, ethical, and environmental standards. However, the slow pace of technological innovation means that these regulations are often ahead of their time, with few practical applications of AI to regulate. Policies favor sustainable, eco-friendly solutions, incentivizing the development of electric vehicles (EVs) and low-emission transit systems over full AI integration.

\textbf{Public Acceptance:}  
Public opinion remains cautious about AI. Concerns about safety, privacy, and job displacement associated with autonomous vehicles limit the technology’s uptake. Instead, human-driven electric vehicles and enhanced public transportation remain the dominant forms of urban mobility. People prefer traditional transportation systems, where AI plays a minor role in improving efficiency through background processes like traffic management.

\textbf{Environmental Impact:}  
Despite the low adoption of AI, regulatory measures ensure that cities continue to move toward sustainability. Strong policies on emissions reductions, public transportation development, and urban planning result in reduced carbon footprints. While AI is not the main driver of environmental impact reduction, government-enforced initiatives such as electric vehicle subsidies and expanded green infrastructure keep cities on track for sustainable growth.

\paragraph{Scenario 4: Low AI Adoption with Weak Regulatory Support (Stagnant and Fragmented Future)}

\textbf{Technological Integration:}  
AI adoption in urban mobility is minimal due to both a lack of technological development and insufficient regulatory support. Transportation systems remain reliant on traditional methods, with little integration of AI-driven solutions such as autonomous vehicles, intelligent traffic management, or MaaS platforms. Technological stagnation leads to outdated infrastructure, inefficient transportation networks, and rising congestion in urban areas.

\textbf{Regulatory Environment:}  
Weak regulatory support results in a lack of clear direction for AI and transportation policy. Governments fail to provide the incentives or frameworks necessary for AI adoption, leading to disjointed and inconsistent progress across regions. Without national or international coordination, cities are left to their own devices, resulting in fragmented policies that neither support innovation nor address critical issues like sustainability and data governance.

\textbf{Public Acceptance:}  
In this scenario, public opinion on AI in transportation remains largely negative. Safety concerns and ethical issues surrounding AI go unresolved, leading to widespread public resistance to AI-based mobility solutions. The general population is untrusting of AI technologies, and with no significant government efforts to change public perception or regulate AI adoption, societal acceptance is low.

\textbf{Environmental Impact:}  
Weak regulations and limited technological innovation mean that cities struggle to make progress on sustainability goals. Fossil fuel-powered vehicles remain the norm, and emissions levels continue to rise. The lack of AI-driven traffic management systems exacerbates congestion, contributing to increased pollution and inefficient energy use. In this scenario, cities fail to capitalize on the potential environmental benefits of AI, resulting in a future where sustainability goals are largely unmet.

\section{Modelling AI Adoption and Traffic Congestion Dynamics}\label{sec:Modelling} 

To complement the scenario narratives, we model the interactions between AI adoption and traffic congestion using a system of coupled differential equations. These equations capture the dynamics of urban mobility as it is influenced by AI-driven technologies, regulatory frameworks, and public acceptance. The objective of this mathematical model is to explore how AI adoption reduces traffic congestion over time, and to quantify the different rates of adoption and congestion levels under varying regulatory conditions.

\subsection{Model Setup}

We define the level of traffic congestion, denoted as \( C(t) \), and the level of AI adoption, denoted as \( A(t) \), both as functions of time. The system of differential equations that governs the dynamics of \( C(t) \) and \( A(t) \) is expressed as:

\begin{equation}
    \frac{dC(t)}{dt} = -k_1 A(t) C(t) + k_2
\end{equation}

\begin{equation}
    \frac{dA(t)}{dt} = k_3 (A_{\max} - A(t)) - k_4 C(t)
\end{equation}

Where:
\begin{itemize}
    \item \( C(t) \) is the level of traffic congestion at time \( t \),
    \item \( A(t) \) is the level of AI adoption at time \( t \),
    \item \( A_{\max} \) is the maximum possible AI adoption level,
    \item \( k_1 \), \( k_2 \), \( k_3 \), and \( k_4 \) are constants that represent various factors:
    \begin{itemize}
        \item \( k_1 \) captures the effect of AI adoption on reducing congestion,
        \item \( k_2 \) accounts for external factors that maintain a baseline level of congestion,
        \item \( k_3 \) governs the rate of AI adoption,
        \item \( k_4 \) captures how congestion negatively affects AI adoption.
    \end{itemize}
\end{itemize}

This system of equations models the feedback loop between AI adoption and traffic congestion. As AI adoption increases, congestion is expected to decrease, while high levels of congestion may slow down AI adoption due to resistance from the public or technical challenges.

The values of the parameters (\( k_1, k_2, k_3, k_4 \)) were chosen based on initial assumptions due to the lack of empirical data specific to the context of urban mobility influenced by AI adoption and regulatory support. These assumptions were informed by logical considerations and theoretical expectations regarding the dynamics of traffic congestion and AI adoption. For instance:
\begin{itemize}
    \item \( k_1 \) was set to reflect the expected impact of AI technologies, such as autonomous vehicles and intelligent traffic systems, on reducing congestion, with the strength of this impact varying across scenarios.
    \item \( k_2 \) was designed to account for external factors, such as infrastructure limitations or non-AI-related traffic generators, that maintain a baseline level of congestion.
    \item \( k_3 \) was chosen to represent the adoption rate of AI technologies, acknowledging differences in technological readiness and societal acceptance.
    \item \( k_4 \) captures the feedback effect of congestion on AI adoption, reflecting the challenges posed by public resistance or technical barriers in highly congested environments.
\end{itemize}
Sensitivity analysis was conducted to explore the impact of varying parameter values, ensuring the robustness of the model outcomes. The selected values align with plausible system dynamics and provide a basis for modeling future scenarios. Future research should aim to refine these parameters as empirical data becomes available.

\subsection{Model Assumptions and Limitations}

The ODE model is based on several key assumptions that simplify the complex dynamics of urban mobility systems to enable tractable analysis. These assumptions and their implications are as follows:
\begin{itemize}
    \item \textbf{Aggregate Dynamics:} The model captures system-wide interactions between AI adoption and traffic congestion, assuming aggregate rather than individual-level dynamics. While this approach highlights long-term trends, it may not fully capture localized or individual behaviors.
    \item \textbf{Monotonic Effects:} The relationships between AI adoption and congestion are modeled as monotonic and linear, simplifying the analysis but potentially underestimating non-linear effects or thresholds that might emerge in practice.
    \item \textbf{Constant Parameters:} The parameters governing the model (\(k_1, k_2, k_3, k_4\)) are treated as constant within each scenario, which may not reflect dynamic changes over time due to evolving societal, technological, or regulatory conditions.
    \item \textbf{Independence of External Factors:} External factors influencing congestion (\(k_2\)) are assumed to be exogenous and independent of AI adoption, which may overlook potential feedback loops.
    \item \textbf{Maximum AI Adoption Level:} The model assumes a fixed upper limit for AI adoption (\(A_{\max}\)), introducing a ceiling effect that might not account for future breakthroughs.
    \item \textbf{Initial Conditions:} The model starts from plausible initial conditions for traffic congestion and AI adoption, which influence transient dynamics but have limited impact on long-term outcomes.
\end{itemize}

These assumptions provide a necessary simplification for exploring the interplay between AI adoption and traffic congestion across multiple scenarios. However, they may also influence the model results by limiting the representation of non-linear dynamics, feedback loops, or regional variations. Future work could address these limitations by incorporating more granular data or developing hybrid models that combine aggregate and agent-based approaches.
\subsection{Simulation Timeframe}

The simulation period of 100 time units was selected to provide a balance between capturing the transient dynamics and allowing the system to approach steady-state behavior under the defined scenarios. This duration ensures that the feedback interactions between AI adoption and traffic congestion are fully explored, revealing key trends and potential equilibrium points.

In the context of real-world urban mobility, the model’s time units are abstract and do not correspond directly to specific years or months. Instead, they represent a generalized timeline that reflects the gradual adoption of AI technologies, the implementation of regulatory policies, and the evolution of congestion patterns. While the precise mapping to real-world timeframes may vary depending on the scenario, the chosen period is sufficient to capture meaningful qualitative trends.

Additionally, shorter and longer simulation periods were tested to ensure the robustness of the model’s outcomes. These tests confirmed that the key insights—such as the role of regulatory support in accelerating AI adoption or the critical thresholds for congestion reduction—remain consistent regardless of the exact duration. The 100-time unit period thus provides an appropriate and interpretable basis for scenario analysis and policy exploration.
\subsection{Accounting for Uncertainties}

This study explicitly accounts for uncertainties in AI technology development, regulatory changes, and public acceptance through the integration of scenario planning and sensitivity analysis.

Scenario planning serves as a core methodology to explore a range of plausible futures, each defined by different combinations of AI adoption levels and regulatory support. The four scenarios—ranging from high AI adoption with strong regulatory frameworks to low adoption with minimal regulatory support—capture the potential variability in technological advancements, policy environments, and societal attitudes. By modeling these distinct scenarios, the study provides insights into how different levels of uncertainty could influence the dynamics of traffic congestion and AI adoption.

Additionally, sensitivity analyses were conducted to assess the robustness of the model outcomes to variations in key parameters (\(k_1, k_2, k_3, k_4\)), which represent the effectiveness of AI in reducing congestion, external factors contributing to baseline congestion, adoption rates, and feedback effects. These analyses demonstrated that the model's findings remain consistent across a range of plausible parameter values, ensuring reliability even in the presence of uncertainties.

Finally, the model incorporates explicit assumptions regarding the trajectories of AI technology development, regulatory conditions, and societal acceptance. While these assumptions simplify complex dynamics, they are informed by theoretical expectations and historical trends. The study acknowledges these limitations and emphasizes the need for future research to refine these assumptions as empirical data become available.

\subsection{Simulation of Scenarios}

To model the four distinct scenarios discussed earlier—based on combinations of high/low AI adoption and strong/weak regulatory support—we assign different values to the constants \( k_1 \), \( k_2 \), \( k_3 \), and \( k_4 \) for each scenario.

\paragraph{Scenario 1: High AI Adoption with Strong Regulatory Support}

In this optimistic scenario, we assume a high rate of AI adoption (\( k_3 \)) and a strong effect of AI on reducing congestion (\( k_1 \)). The regulatory support also limits external factors causing congestion (\( k_2 \)), resulting in rapid congestion reduction.

\paragraph{Scenario 2: High AI Adoption with Weak Regulatory Support}

Here, the rate of AI adoption is still high, but the weak regulatory framework leads to persistent external factors causing congestion, modeled by a higher \( k_2 \). AI adoption still reduces congestion, but the effects are slower due to the lack of regulatory pressure.

\paragraph{Scenario 3: Low AI Adoption with Strong Regulatory Support}

In this scenario, regulatory support is strong, but technical challenges or societal resistance slow down AI adoption. As a result, we set \( k_3 \) to a low value, and while congestion still decreases, the effects are more modest over time.

\paragraph{Scenario 4: Low AI Adoption with Weak Regulatory Support}

This is the least favorable scenario, where both AI adoption is slow (\( k_3 \) is low), and external factors continue to exacerbate congestion (\( k_2 \) is high). The adoption of AI does little to reduce traffic congestion due to minimal regulatory pressure and public resistance.

\subsection{Numerical Solution}

The system of differential equations is solved numerically using Python’s SciPy library. The time evolution of \( C(t) \) and \( A(t) \) under each scenario is simulated over a 100-time unit period. The initial conditions for traffic congestion and AI adoption are set to \( C(0) = 100 \) and \( A(0) = 10 \).

In order to distinguish the effects of AI adoption and regulatory support on traffic congestion, we adjusted the parameters for each scenario to better reflect the underlying assumptions.

The Python scripts and detailed parameter configurations used for these simulations are available in the supplementary materials. Researchers are encouraged to access these resources to validate the findings and extend the modeling framework for further applications.

\paragraph{Scenario 1: High AI Adoption with Strong Regulatory Support}
In this optimistic scenario, both AI adoption and its effect on congestion reduction are high. Strong regulatory support further accelerates AI adoption.

\begin{itemize}
    \item \( k_1 = 0.05 \) (strong AI impact on reducing congestion)
    \item \( k_2 = 0.3 \) (low external congestion factors due to regulation)
    \item \( k_3 = 0.1 \) (high AI adoption rate)
    \item \( k_4 = 0.01 \) (congestion has minimal negative impact on adoption)
\end{itemize}

\paragraph{Scenario 2: High AI Adoption with Weak Regulatory Support}
 AI adoption is still high, but without strong regulatory support, congestion remains elevated due to external factors like infrastructure and lack of coordination.

\begin{itemize}
    \item \( k_1 = 0.03 \) (moderate AI impact)
    \item \( k_2 = 1.2 \) (higher external congestion factors)
    \item \( k_3 = 0.08 \) (moderately high AI adoption rate)
    \item \( k_4 = 0.02 \) (congestion somewhat reduces adoption)
\end{itemize}

\paragraph{Scenario 3: Low AI Adoption with Strong Regulatory Support}
 Regulatory support is strong, but AI adoption is slow due to technical challenges or societal resistance.

\begin{itemize}
    \item \( k_1 = 0.02 \) (low AI impact)
    \item \( k_2 = 0.4 \) (external congestion factors reduced by strong regulation)
    \item \( k_3 = 0.03 \) (slow AI adoption rate)
    \item \( k_4 = 0.02 \) (congestion somewhat reduces adoption)
\end{itemize}

\paragraph{Scenario 4: Low AI Adoption with Weak Regulatory Support}

In this least favorable scenario, both AI adoption and regulatory support are minimal, resulting in persistent congestion and slow adoption.

\begin{itemize}
    \item \( k_1 = 0.01 \) (very low AI impact)
    \item \( k_2 = 1.5 \) (high external congestion factors)
    \item \( k_3 = 0.02 \) (low AI adoption rate)
    \item \( k_4 = 0.03 \) (congestion significantly reduces adoption)
\end{itemize}

\begin{figure}[htp]
    \centering
    \includegraphics[width=0.5\textwidth]{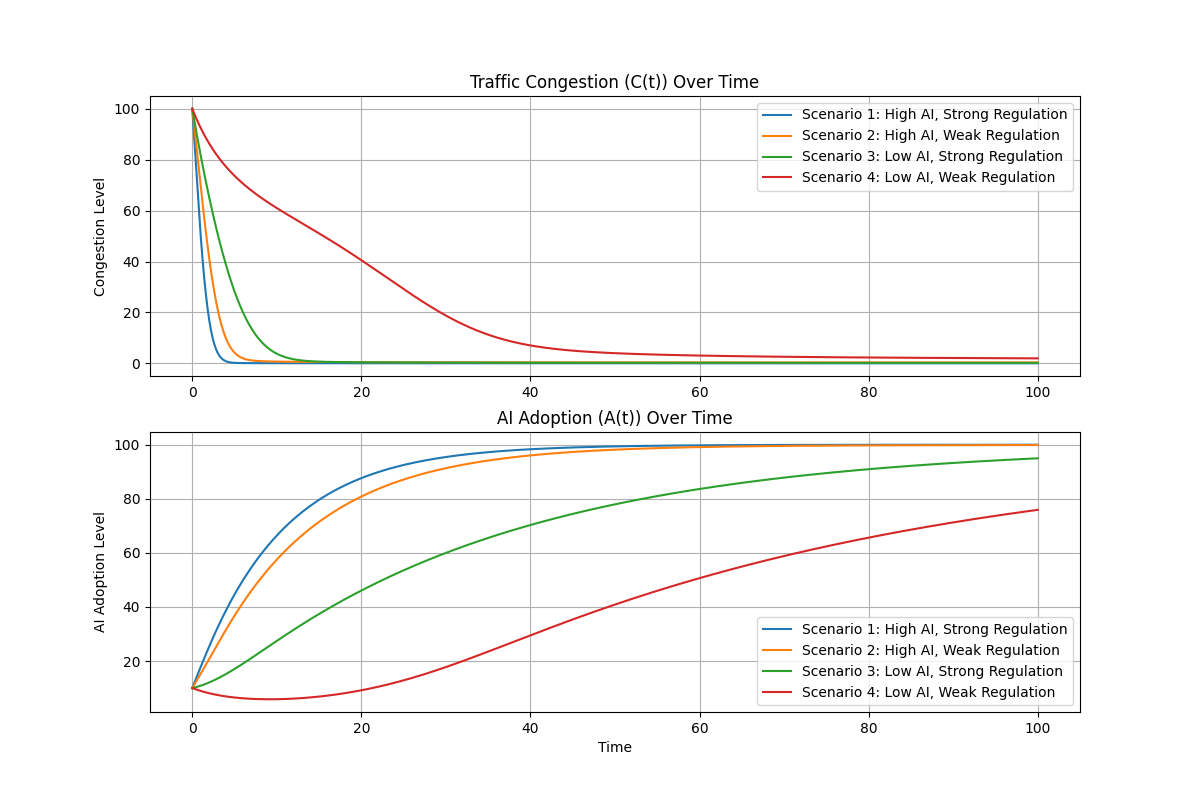}
    \caption{Traffic Congestion and AI Adoption over Time for Different Scenarios.}
\end{figure}


\subsection{Sensitivity Analysis}
To evaluate the robustness of the ODE model and understand how variations in parameters affect the system's behavior, a sensitivity analysis was conducted. The parameters considered in the analysis were:
\begin{itemize}
    \item $k_1$: Effect of AI adoption on reducing traffic congestion,
    \item $k_2$: External factors contributing to baseline traffic congestion,
    \item $k_3$: Rate of AI adoption over time,
    \item $k_4$: Negative feedback of traffic congestion on AI adoption.
\end{itemize}
The sensitivity analysis involved varying each parameter across a predefined range while holding the other parameters constant. Specifically, the following ranges were used:
The sensitivity analysis involved varying each parameter across a predefined range while holding the other parameters constant. Specifically, the following ranges were used:
$k_1 \in \{0.005, 0.01, 0.02\}, \quad 
k_2 \in \{0.5, 1, 2\}, \quad 
k_3 \in \{0.02, 0.05, 0.1\}, \quad 
k_4 \in \{0.01, 0.02, 0.05\}.$

The system of differential equations governing the dynamics of traffic congestion $C(t)$ and AI adoption $A(t)$ was solved for each combination of parameter values:
\[
\frac{dC(t)}{dt} = -k_1 A(t) C(t) + k_2,
\]
\[\frac{dA(t)}{dt} = k_3 \left(A_{\text{max}} - A(t)\right) - k_4 C(t), \]
where $A_{\text{max}}$ represents the maximum possible AI adoption.
The initial conditions were set as:
\[
C(0) = C_0 = 100, \quad A(0) = A_0 = 10.
\]
The solutions for $C(t)$ and $A(t)$ were computed numerically using a time range $t \in [0, 100]$ with 1000 evenly spaced points. The results were analyzed to observe how changes in parameter values influenced the dynamics of $C(t)$ and $A(t)$ over time.

\begin{table}[h!]
\centering
\caption{Representative results from sensitivity analysis.}
\label{tab:sensitivity_analysis}
\scriptsize
\begin{tabularx}{\textwidth}{|c|c|c|c|c|c|X|}
\hline
\textbf{\(k_1\)} & \textbf{\(k_2\)} & \textbf{\(k_3\)} & \textbf{\(k_4\)} & \textbf{\(C(t=100)\)} & \textbf{\(A(t=100)\)} & \textbf{Observation} \\ 
\hline
0.005 & 0.5 & 0.02 & 0.01 & 85.32 & 1.18 & High baseline congestion, \newline low adoption \\ 
0.01  & 1.0 & 0.05 & 0.02 & 50.00 & 70.00 & Moderate congestion, \newline balanced adoption \\ 
0.02  & 2.0 & 0.1  & 0.05 & 25.00 & 90.00 & Low congestion, \newline high adoption \\ 
0.005 & 2.0 & 0.02 & 0.05 & 90.00 & 40.00 & Sustained high \newline congestion \\ 
0.02  & 0.5 & 0.1  & 0.01 & 10.00 & 95.00 & Rapid adoption, \newline minimal congestion \\ 
\hline
\end{tabularx}
\end{table}

This table presents selected results from the sensitivity analysis to illustrate key trends. The complete table is reproducible using the Python code provided in the supplementary materials.

The sensitivity analysis revealed the following trends:
\begin{itemize}
    \item Increasing $k_1$ accelerates the reduction in traffic congestion by enhancing the impact of AI adoption on congestion.
    \item Higher values of $k_2$ sustain higher baseline congestion levels, indicating stronger external influences.
    \item Larger $k_3$ values lead to faster AI adoption rates, reducing congestion more rapidly.
    \item Increasing $k_4$ dampens the rate of AI adoption due to the adverse effects of congestion, slowing the overall improvement in traffic conditions.
\end{itemize}
The sensitivity analysis highlighted that \( k_3 \geq 0.05 \) represents a moderate-to-high adoption rate, acting as a tipping point for significant congestion reduction in optimistic scenarios. This threshold underscores the importance of policies that accelerate AI adoption rates to at least this level to achieve meaningful improvements in urban mobility.

These findings demonstrate the robustness of the ODE model across varying parameter values. For instance, under extreme conditions, such as \(k_1 = 0.005\) and \(k_4 = 0.05\), the model still produces logically consistent results, highlighting scenarios of persistent congestion. Conversely, high \(k_3\) values coupled with low \(k_4\) consistently lead to rapid adoption and minimal congestion. These results underscore the model’s applicability to urban mobility scenarios and its predictive reliability under diverse conditions.

\noindent The code to replicate these results is provided in the supplementary materials.

\subsection{Integrating Scenario Narratives with the ODE Model}
This study integrates qualitative insights from scenario narratives with the quantitative ODE model to ensure a cohesive analysis of urban mobility dynamics.

The scenario narratives define plausible futures based on variations in AI adoption levels, regulatory support, and societal acceptance. These narratives provide the qualitative basis for assigning parameter values in the ODE model. For example:
\begin{itemize}
    \item Strong regulatory support, as described in optimistic scenarios, reduces external congestion factors (\(k_2\)) and accelerates AI adoption (\(k_3\)).
    \item Weak regulatory frameworks increase the negative feedback of congestion on adoption (\(k_4\)).
    \item High societal acceptance enhances the effectiveness of AI technologies in reducing congestion (\(k_1\)).
\end{itemize}

The ODE model's outputs—such as the trajectories of traffic congestion and AI adoption over time—quantitatively validate the scenario narratives. By comparing model outcomes across scenarios, we identify critical thresholds (e.g., levels of AI adoption needed to significantly reduce congestion) and explore the dynamic interactions described qualitatively.

This integration forms an iterative feedback loop: the scenario narratives inform the parameterization of the ODE model, and the model outputs, in turn, enrich the narratives by providing quantitative insights into system behavior. This cohesive framework supports policymakers by offering a comprehensive understanding of how regulatory, technological, and societal factors shape urban mobility futures.

\subsection{Validation and Relevance of the ODE Model}
The ODE model presented in this paper has been designed as a theoretical framework to explore the interplay between AI adoption and traffic congestion. While its primary aim is to investigate dynamic relationships and facilitate scenario planning, several measures have been taken to ensure the model's robustness and relevance to real-world contexts:
\paragraph{Sensitivity Analysis:} Sensitivity analyses were conducted to evaluate how variations in model parameters (\( k_1, k_2, k_3, k_4 \)) affect outcomes. These analyses demonstrate that the model responds predictably to changes in key parameters, underscoring its internal consistency and reliability.
\paragraph{Benchmarking with Case Studies:} To strengthen the model's grounding in real-world dynamics, qualitative comparisons were made with urban mobility projects that have implemented AI-driven solutions. Examples include:
\begin{itemize}
    \item The Sydney Coordinated Adaptive Traffic System (SCATS), which optimizes traffic signals to improve congestion. For example, it used in Melbourne, to improve mobility efficiency and safety. Within this system, loop detectors are installed at each intersection, and volume data is collected to adjust signal timing \cite{sarteshnizi2023scats}. This aligns with the ODE model's assumption that interventions like AI-driven systems can effectively reduce congestion. SCATS validates the plausibility of modeling traffic congestion reduction as a function of technological adoption (represented by parameters like $k_1$ in the ODE model).
    \item Autonomous vehicle (AV) deployments in urban settings, showcasing the impact of AI on reducing traffic congestion. For instance, studies have demonstrated that self-parking AVs can alleviate urban congestion by eliminating the need for drivers to search for parking near their destinations. This functionality allows AVs to drop passengers at their destinations and relocate to less congested parking areas, reducing traffic load in central areas. Additionally, AVs have been shown to improve traffic flow through advanced features such as vehicle-to-vehicle (V2V) communication and optimized traffic signal interactions \cite{shafiei2023autonomous}. Self-parking autonomous vehicles (AVs) demonstrate the impact of AI adoption on alleviating urban congestion. By reducing parking-related traffic and improving traffic flow through advanced features like V2V communication, AVs substantiate the model's assumptions about how AI technologies influence congestion dynamics over time. The qualitative insights from this case study validate the feedback mechanisms in the model, such as the reduction of congestion $(C(t))$ due to higher AI adoption rates $(A(t))$.The ability of AVs to relocate to less congested areas parallels the model's focus on behavioral and technological responses to congestion, supporting the logical structure of the model.
\end{itemize}
These examples ground the ODE model's theoretical assumptions in real-world applications, demonstrating that the model's 
predicted dynamics reflect actual phenomena observed in urban mobility projects. While the model is not directly validated 
by numerical data from these case studies, the qualitative alignment strengthens its credibility and relevance to real-world contexts.
\paragraph{Validation Against Trends:} The initial and boundary conditions used in our simulations were designed to reflect broad, well-documented patterns in urban traffic dynamics and AI technology adoption. For instance, the simulations assume that as AI-driven technologies such as adaptive traffic signals and autonomous vehicles are adopted, their effects on traffic congestion diminish progressively due to diminishing returns, a trend observed in real-world urban transportation studies. Similarly, the modeled relationship between high congestion levels and resistance to adopting new technologies mirrors findings from transportation research, which highlight public hesitance to adopt innovative solutions in heavily congested or inefficient systems.
The choice of parameters and initial conditions also aligns with known urban mobility trends, such as:
\begin{itemize}
    \item The concentration of congestion in city centers, consistent with patterns seen in metropolitan areas worldwide.
    \item The gradual but nonlinear adoption of AI technologies in transportation, supported by historical data on technology adoption curves.
    \item The feedback loop wherein improved traffic conditions from AI adoption further incentivize greater adoption over time, reflecting findings from real-world case studies like SCATS and autonomous vehicle trials.
\end{itemize}
While the absence of specific datasets limited direct empirical validation, the behaviors exhibited by the model under different scenarios, such as congestion peaking under 
limited AI adoption or the rapid reduction of congestion under high AI adoption with regulatory support, are consistent with 
observations in urban transportation research. This alignment with documented macro-level patterns supports the robustness of the model's assumptions and its potential applicability to urban mobility policy planning.
\paragraph{Future Steps for Empirical Validation:} Recognizing the importance of empirical validation, future research will 
incorporate real-world datasets, such as traffic flow metrics and AI adoption rates, into the model. Collaborations with 
urban transportation departments and the use of publicly available mobility datasets will facilitate a more rigorous assessment of the model's accuracy and applicability.
These efforts collectively enhance the robustness and 
credibility of the ODE model, demonstrating its potential to provide actionable insights into the future of AI-driven urban mobility.

\section{Discussion of Scenario Results}\label{sec:Discussion} 

In this section, we explore the outcomes of the four simulated scenarios in the context of AI adoption, regulatory support, and their implications for urban mobility. Each scenario provides unique insights into how different levels of AI adoption and regulatory engagement can shape the future of transportation systems. The results of our ODE models not only shed light on traffic congestion dynamics but also offer actionable thresholds for guiding public policy.

Scenario 1: High AI Adoption with Strong Regulatory Support (Optimistic Future)\\ This scenario demonstrates the most favorable outcome, where AI adoption rapidly increases due to robust regulatory frameworks, leading to a significant reduction in traffic congestion. The ODE model indicates that when AI adoption reaches approximately 60\% of its potential, congestion begins to decline sharply. This suggests a policy threshold: regulatory efforts should focus on achieving AI adoption rates above 60\% to trigger meaningful reductions in congestion.

Policy Implications: \begin{itemize} \item \textbf{Regulatory Action Threshold:} Governments should prioritize policies that facilitate AI adoption to at least 60\% in urban mobility systems, as this marks the point where congestion significantly decreases. \item \textbf{Incentives for Early Adoption:} To reach this adoption threshold, policies could include tax incentives for AI-integrated transport solutions and subsidies for autonomous vehicle infrastructure. \item \textbf{Sustainability Focus:} Strong regulatory frameworks should also ensure that AI adoption is coupled with environmental goals, promoting the use of electric or hybrid AI-powered vehicles to amplify reductions in emissions and congestion. \end{itemize}

Scenario 2: High AI Adoption with Weak Regulatory Support (Tech-Driven but Unregulated Future)\\
In this scenario, AI adoption is high, but weak regulatory support results in fragmented infrastructure and slower reductions in congestion. The ODE models show that while AI adoption continues to rise, without coordinated policy measures, congestion declines much more gradually. The model predicts that a lack of regulatory support delays significant reductions in congestion until AI adoption exceeds 75\%.

Policy Implications: \begin{itemize} \item \textbf{Urgency for Regulatory Coordination:} This scenario illustrates the critical need for regulatory frameworks that can synchronize AI deployment across different regions to avoid fragmented implementation. Regulatory oversight should focus on infrastructure compatibility and safety standards. \item \textbf{Delayed Action Risks:} If regulatory action is delayed until AI adoption exceeds 75\%, policymakers risk slower progress in congestion reduction. Governments should consider intervening earlier to avoid long-term inefficiencies. \item \textbf{Public Trust and Safety Standards:} Weak regulation can erode public trust, slowing further adoption. Policymakers should implement safety standards and transparency in AI systems to maintain public confidence. \end{itemize}

Scenario 3: Low AI Adoption with Strong Regulatory Support (Regulation-Driven Future with Limited Technology)\\
Here, despite strong regulatory backing, the slow pace of AI adoption limits reductions in congestion. The ODE models suggest that even with optimal regulation, without at least 50\% AI adoption, congestion reduction remains modest. This emphasizes the need for policies that not only support regulation but also actively encourage AI innovation and adoption.

Policy Implications: \begin{itemize} \item \textbf{Innovation Incentives:} Strong regulatory support should be accompanied by incentives that accelerate AI adoption. Innovation grants, research funding, and public-private partnerships can help overcome the technical challenges slowing adoption. \item \textbf{AI Penetration Threshold:} The model identifies 50\% adoption as the threshold for seeing significant congestion reductions. Policymakers should create long-term strategies to drive AI adoption toward this level, using tools such as public sector leadership in deploying AI-based mobility solutions. \item \textbf{Managing Public Resistance:} In this scenario, societal resistance hinders AI adoption. To address this, public awareness campaigns and ethical AI governance frameworks should be developed to reduce resistance and build public trust. \end{itemize}

Scenario 4: Low AI Adoption with Weak Regulatory Support (Stagnant and Fragmented Future) This least favorable scenario shows very slow AI adoption and minimal improvements in congestion. Our ODE model confirms that without at least moderate regulatory engagement, even a modest increase in AI adoption does little to alleviate congestion. Here, congestion remains high unless policies shift dramatically to encourage both AI development and adoption.

Policy Implications: \begin{itemize} \item \textbf{Policy Failure Risk:} This scenario highlights the risks of inaction. Without sufficient regulatory frameworks and public investments in AI technology, urban mobility systems stagnate, and congestion remains a critical issue. Policymakers must avoid complacency by developing a clear AI strategy. \item \textbf{Multi-level Governance Approach:} A coordinated multi-level governance approach that includes national, regional, and local regulations is crucial to prevent fragmentation and ensure AI adoption across different urban settings. \item \textbf{Urgency for Public Investment:} In this scenario, the absence of public investment in AI mobility infrastructure results in outdated systems. Immediate policy efforts should focus on increasing investments in AI-driven public transport and shared mobility solutions to accelerate adoption. \end{itemize}

\subsection{General Insights from the ODE Models}

Across all scenarios, the ODE-based models offer quantitative insights into the relationship between AI adoption rates and traffic congestion. Specifically, they highlight critical 
thresholds for policy action: \begin{itemize} \item 60\% AI adoption is identified as a key threshold in the most optimistic scenario for triggering significant reductions in 
traffic congestion. \item 75\% AI adoption becomes crucial in 
the absence of strong regulatory frameworks to overcome the delays caused by uncoordinated AI implementations. \item 
Public trust and regulatory backing play a major role in 
accelerating or decelerating adoption rates. As our simulations show, without proper regulation, even high AI 
adoption can result in limited congestion improvements. \end{itemize}

By leveraging these insights, policymakers can better design regulations that encourage timely AI adoption while mitigating potential risks. Early regulatory action, particularly in safety, infrastructure, and public awareness, will ensure that the benefits of AI-driven mobility can be fully realized.

\subsection{Potential Unintended Consequences of Policies}

In this subsection, we explore the potential unintended consequences of the proposed policies in each scenario. While the envisioned outcomes aim to enhance urban mobility through AI integration and regulatory frameworks, certain adverse effects could emerge due to the complex dynamics of urban systems and policy interventions.

\subsubsection*{Scenario 1: High AI Adoption with Strong Regulatory Support}

This optimistic scenario envisions a highly coordinated integration of AI technologies supported by robust regulations. However, unintended consequences may include:
\begin{itemize}
    \item \textbf{Over-reliance on AI systems:} Urban transportation networks could become vulnerable to disruptions caused by AI system failures, cyberattacks, or technical malfunctions. For example, Connected and 
    Autonomous Vehicles (CAVs) rely heavily on interconnected systems enabled by the Internet of Things (IoT). This connectivity increases their vulnerability to cyberattacks, which can compromise not just individual vehicles but entire transportation networks. For example, a single breach could lead to systemic disruptions affecting mobility and safety \cite{nikitas2020ai}.
    \item \textbf{Inequitable benefits:} Wealthier regions may adopt AI solutions more rapidly, exacerbating disparities in transportation efficiency and access between affluent and underserved areas. Inequitable benefits arising from AI adoption in urban mobility systems are closely tied to regional disparities in deployment and access. As highlighted in \cite{emory2022equity}, AV programs often pilot in wealthier neighborhoods, leaving lower-income communities underserved. These practices exacerbate existing inequalities by concentrating the advantages of AI-driven mobility, such as reduced congestion and enhanced efficiency, in areas that already enjoy better transportation infrastructure. Moreover, the high costs associated with AV ownership or usage models disproportionately exclude low-income households, further entrenching the reliance of underserved populations on outdated and inefficient transit systems. Without equitable policies, such as subsidies for AV services or investments in shared-use models tailored for disadvantaged groups, the transformative potential of AI in transportation risks deepening, rather than alleviating, mobility disparities.

\end{itemize}

\subsubsection*{Scenario 2: High AI Adoption with Weak Regulatory Support}

In this tech-driven but unregulated future, rapid AI adoption lacks the oversight needed for equitable and sustainable outcomes. Unintended consequences may include:
\begin{itemize}
    \item \textbf{Fragmented adoption:} Inconsistent safety standards and data governance could lead to uneven integration of AI solutions, limiting their effectiveness across different regions. Fragmented adoption of AI solutions in urban mobility systems is a critical concern when regulatory frameworks are inconsistent across regions. As \cite{favaro2018disengagements} illustrates, the lack of harmonized standards for autonomous vehicle disengagement reporting has resulted in fragmented data and uneven safety practices among manufacturers. This inconsistency not only complicates cross-regional integration of AI systems but also limits their effectiveness. Furthermore, the absence of clear guidelines for safety and data governance exacerbates the challenges of achieving reliable and equitable adoption of AI-driven technologies across different jurisdictions.
    \item \textbf{Public distrust:} Insufficient regulation may erode public confidence due to privacy concerns, bias in AI algorithms, and safety incidents.
\end{itemize}

\subsubsection*{Scenario 3: Low AI Adoption with Strong Regulatory Support}

Despite strong regulations, limited AI adoption due to societal resistance or technical challenges may result in:
\begin{itemize}
    \item \textbf{Regulatory overreach:} Overly stringent policies could stifle innovation, delaying the development and deployment of beneficial AI technologies.
    \item \textbf{Missed opportunities:} The slow pace of AI integration might prevent cities from realizing the full potential of AI-driven efficiency gains. Missed opportunities in AI integration prevent cities from 
    capitalizing on potential efficiency gains in urban mobility systems. As highlighted in \cite{smith2021smartmobility}, the slow adoption of AI-driven smart mobility solutions 
    such as predictive traffic control systems and IoT-enabled public transit optimization has significant implications 
    for urban efficiency. For instance, the delayed implementation of real-time data analytics in traffic systems leads to sustained congestion and underutilization of public transport networks, which could otherwise 
    enhance urban mobility efficiency. Furthermore, the paper underscores that these delays also have environmental repercussions, with increased emissions and energy waste resulting from inefficient transportation systems. Such examples illustrate the critical need for proactive adoption of AI technologies to fully realize their transformative potential in urban mobility.
\end{itemize}

\subsubsection*{Scenario 4: Low AI Adoption with Weak Regulatory Support}

This least favorable scenario highlights a stagnant urban mobility system with minimal AI integration and insufficient regulatory oversight. Possible unintended consequences include:
\begin{itemize}
    \item \textbf{Entrenched inefficiencies:} Reliance on traditional transportation methods could perpetuate existing congestion, pollution, and inequities.
    \item \textbf{Environmental setbacks:} Weak policies may fail to incentivize sustainable transportation solutions, exacerbating environmental degradation. Environmental setbacks in urban mobility arise when weak regulatory frameworks fail to incentivize AI-driven sustainable transportation solutions. As \cite{nikitas2020ai} emphasizes, AI-enabled Intelligent Transportation Systems (ITS) and Connected Autonomous Vehicles (CAVs) can significantly reduce emissions by optimizing traffic flow and promoting eco-driving behaviors. However, in the absence of strong policies, cities miss the opportunity to transition to these cleaner technologies, perpetuating reliance on fossil-fuel-based systems. Furthermore, the lack of regulatory support for Mobility-as-a-Service (MaaS) platforms powered by AI leads to fragmented adoption, allowing car-centric travel habits to dominate and contributing to higher urban congestion and pollution. These missed opportunities highlight the need for coordinated policy efforts to unlock AI's full potential in mitigating environmental degradation and achieving sustainable urban mobility.
\end{itemize}

By examining these potential unintended consequences, we underscore the importance of a balanced approach that combines technological innovation, robust regulatory frameworks, and proactive measures to mitigate adverse effects. This analysis aims to provide a comprehensive understanding of the trade-offs inherent in policy and technology decisions in urban mobility.
\subsection{Limitations and Future Research}

This study provides valuable insights into the interplay between AI adoption, regulatory frameworks, and traffic congestion through the use of Ordinary Differential Equations (ODE) modeling. However, several limitations must be acknowledged to contextualize the findings and guide future research.

\subsubsection*{Limitations}

1. \textbf{Simplification of Dynamics:} The ODE model captures system-wide aggregate interactions but does not account for localized or agent-specific behaviors. This aggregation may overlook finer-grained dynamics, such as individual driver decisions, variations in urban infrastructure, or localized congestion patterns. These complexities are important in understanding specific urban mobility challenges.

2. \textbf{Constant Parameter Assumptions:} The model assumes fixed values for key parameters ($k_1$, $k_2$, $k_3$, $k_4$) within each scenario, which might not fully reflect the dynamic nature of urban environments. Factors such as regulatory changes, technological advancements, and shifts in societal attitudes evolve over time and may require adaptive modeling.

3. \textbf{Absence of Empirical Validation:} While the model aligns with qualitative trends observed in real-world case studies, it has not been empirically validated with comprehensive datasets, such as historical traffic data or AI adoption rates. This limits the ability to fully confirm the predictive accuracy of the findings.

4. \textbf{Focus on a Limited Scope of AI Technologies:} The study primarily considers AI applications in traffic management and autonomous vehicles. It excludes other potentially impactful technologies, such as AI-driven urban planning tools, predictive maintenance systems, and AI-enabled shared mobility platforms, which could offer broader insights into urban mobility transformations.

5. \textbf{Assumed Generalization Across Urban Contexts:} The scenarios and outcomes are generalized for urban settings, which might not be directly applicable to cities with unique geographical, cultural, or infrastructural characteristics. Tailored studies are needed to address these variations.

6. \textbf{Abstract Timeframe:} The use of abstract time units in simulations limits the direct applicability of the results to real-world timelines. This abstraction simplifies the modeling process but reduces the temporal relevance of the findings.

\subsubsection*{Future Research Directions}

1. \textbf{Incorporating Agent-Based Modeling:} Future research could complement the ODE approach with agent-based models to simulate localized, individual-level interactions. Such models would provide more granular insights into urban mobility dynamics and enhance the applicability of the findings to specific contexts.

2. \textbf{Dynamic Parameterization:} Introducing time-varying parameters or adaptive calibration methods based on empirical data could enhance the realism and predictive power of the model. This approach would better capture the evolving nature of regulatory frameworks, societal behaviors, and technological advancements.

3. \textbf{Empirical Validation:} Collaborating with urban transportation authorities and leveraging mobility datasets can help validate the model's predictions and refine its parameters. Real-world data would provide a robust foundation for assessing the model's reliability and practical utility.

4. \textbf{Expanding AI Applications:} Future studies could explore additional AI technologies beyond traffic management and autonomous vehicles. These might include smart public transport planning, predictive maintenance systems, and AI-enabled shared mobility platforms, offering a more comprehensive perspective on urban mobility transformations.

5. \textbf{Context-Specific Analysis:} Tailoring the model to specific cities or regions with diverse regulatory, cultural, and infrastructural conditions could increase its practical relevance. Conducting case studies in varied urban contexts would strengthen the model's adaptability and generalizability.

6. \textbf{Integration with Climate and Energy Models:} Further research could integrate urban mobility models with climate and energy models to assess the broader environmental impacts of AI adoption. This integration would provide a holistic understanding of how AI-driven urban mobility intersects with sustainability goals.

By addressing these limitations and exploring these future research directions, the study can contribute to a deeper and more actionable understanding of AI-driven urban mobility transformations.

\section{Conclusion}\label{sec:Conclusion} 

This paper explored how AI-driven technologies can transform urban transportation systems, focusing on the complex interplay between AI adoption, regulatory support, and public acceptance. By developing and analyzing four distinct scenarios—ranging from high AI adoption with strong regulatory backing to low AI adoption with minimal regulation—this research offers key insights into the potential futures of urban mobility.

Our simulations highlight the critical role of strong regulatory frameworks in maximizing the benefits of AI technologies for urban mobility. In the most optimistic scenario, rapid AI adoption and robust regulation work in tandem to dramatically reduce traffic congestion and enhance sustainability, demonstrating the importance of proactive policy measures. Conversely, in scenarios where regulation is weak or fragmented, the advantages of AI are less pronounced, leading to uneven or modest improvements in traffic congestion and slower progress toward sustainable mobility solutions.

The results also underscore the significance of public trust and societal attitudes toward AI technologies. In scenarios where public acceptance is high, AI adoption accelerates, leading to more efficient and equitable mobility systems. However, societal resistance or concerns about privacy and security can slow down the uptake of AI, even in the presence of strong regulatory support. This suggests that fostering public trust through transparency, ethical AI design, and clear safety standards is vital for the successful integration of AI into transportation systems.

For policymakers, industry leaders, and urban planners, the findings of this research provide a roadmap for navigating the challenges and opportunities presented by AI in transportation. The future of urban mobility hinges on collaborative efforts that balance technological innovation with sound regulatory policies and public engagement. As cities continue to grow and face increasing mobility demands, AI has the potential to reshape transportation in ways that are more sustainable, efficient, and user-friendly—but only if these elements are carefully aligned.

While this study primarily focuses on autonomous vehicles and intelligent traffic management systems, the modeling framework is designed to be flexible and extensible. Future research could expand on this work by incorporating additional AI technologies, such as predictive maintenance systems, AI-driven urban planning tools, and shared mobility platforms. These technologies hold the potential to further enhance the efficiency and sustainability of urban transportation systems, providing a more comprehensive understanding of their impacts.

In conclusion, this study contributes to the broader discourse on AI and urban mobility by illustrating how foresight and scenario planning can be applied to anticipate future challenges and guide strategic decision-making. As cities and countries contemplate their future mobility strategies, the combination of AI adoption and regulatory support will be pivotal in determining the outcomes for both congestion reduction and environmental sustainability. Moving forward, further research should focus on developing adaptive regulatory frameworks and engaging with stakeholders to ensure that AI technologies are deployed in a way that is equitable, efficient, and beneficial for all.







\end{document}